\documentclass[preprint,aps]{revtex4}
\usepackage{graphicx}
\usepackage{float}
\usepackage{amsmath}    

\begin{document}
\title{OPERA neutrinos and superluminal helical motion}
\author{E. Canessa\footnote{e-mail: canessae@gmail.com}}
\affiliation{Science \& Technology Collaborium, Italy
\vspace{3cm}
}

\begin{abstract}
We pinpoint how a subatomic particle with non-zero mass may attain, 
in principle, velocities faster-than-light by travelling in helical 
motion in the limit of very large momentum.  This is an educated 
guess by virtue of the MINOS and OPERA experiments on eventual 
superluminal propagation of neutrinos.  

\vspace{1.0cm}


\end{abstract}
\maketitle

Our main focus in this brief note is to pinpoint the possibility of having 
velocities faster-than-light ($v>c$) when considering helical motion for 
particles of non-zero mass $m$ in the limit of very large momentum ($mv$).  
This is an educated guess to provide a side view regarding eventual 
superluminal neutrinos 
after MINOS \cite{Min11}  and OPERA \cite{Ope11} experiments.  Alternative 
discussions include the assumption of tachyons \cite{Fey67,Ale11},
neutrino dark energy \cite{Ciu11}, 
environmental fifth-force hypothesis \cite{Ior11}, possible statistical mechanisms 
\cite{Ali11,Gon11} and Lorentz symmetry violations \cite{Kli11} to mention a few.  
The tiny influence of Coriolis effect in neutrino speed measurements between
CERN and the Gran Sasso Lab is discussed in \cite{Kuh11}.
We introduce next our bold assumptions on the theme.

Elementary subatomic neutrinos are electrically neutral, weakly interacting, and have
a small but non-zero mass. Being electrically neutral, they are able to pass through 
ordinary matter being 'affected' as shown below.  To begin, we study first an energy-dependent 
speed model following discussions in \cite{Keh11,Ame11}.

The total energy of fast moving objects satisfies the key Einstein equation
$E =  m c^{2}$.  In terms of the momentun $p = mv = m_{o}v/\sqrt{1 - (v/c)^{2}}$, 
this relation can be rewritten as 
\begin{equation}\label{eq:Ein} 
E^{2} =  m_{o}^{2} c^{4} + c^{2} p^{2} \;\;\; , 
\end{equation}
with $m_{o}$ the object rest mass.  If $p$ is very large so that
$m_{o}^{2} c^{4} << c^{2} p^{2}$ (which means $v >> c/\sqrt{2}$ or velocities $\sim 0.7 c$ 
and greater), then the derivative of the above energy-momentum formula can be approximated as 
\begin{equation}
dE \approx c\; dp = c \; d(mv)\;\;\; . 
\end{equation}
This amount of energy (work) may be seen as transferred by a force $F$ acting 
through a distance $r$ relating Newton's second law of motion $F = dp/dt = (dp/dr) v$,
with velocity $v = dr/dt$.  Thus, it follows a 'new class' of Newtonian dynamics such 
that
\begin{equation}
F_{N} =  \frac{dE}{dr} \approx \left( \frac{c}{v} \right) \; \frac{d}{dt}(mv)  =  mc \; \left( \frac{dv}{dr} \right) \;\;\; .
\end{equation}

Let us consider next that the direction of motion is constantly changing through a curved
path (say, circular in 2D and helix in 3D), then the particle exerts an acceleration 
associated to the outward centrifugal force expressed as
\begin{equation}\label{eq:Cen} 
F_{c} = - m \; \left( \frac{v^{2}}{r} \right) \;\;\; .
\end{equation}
In deriving this force, one approximates the {\it arc} of the travelling curved
path as $S = r\; \theta$, 
with $\theta$ the angle measured with respect to the center of curvature and the
radial direction.  On the other hand, another path can similarly be drawn for the 
velocities such that $\Delta v = v\; \theta$.  In the limit of small $\theta > 0$,
or small gradients $\Delta$ (with $S \rightarrow \Delta S$), these two relations 
lead to
\begin{equation}\label{eq:rv} 
\frac{dv}{v}  \approx \frac{dS}{r} \;\;\; .
\end{equation}
Therefore for the closed dynamical system to be in equilibrium the sum of forces 
$F_{N} + F_{c} \equiv 0$, so after some little algebra we have
\begin{equation}\label{eq:vc} 
 \frac{v}{c} \approx  \frac{r}{v} \; \left( \frac{dv}{dr} \right) = \frac{dS}{dr} \;\;\; ,
\end{equation}
which implies
\begin{equation}\label{eq:spiralicity} 
 \frac{v}{c} \approx \theta \;\;\; .
\end{equation}
The special relativity derivation used in this approximation implies $\theta <1$ 
to avoid infinite energies from $E = mc^{2}$, but it is still illustrative to pursue 
research MINOS and OPERA results.  

The motion of neutrinos as described by the Dirac equation with
a neutrino mass matrix \cite{Wid96}, and a Lorentz scalar imaginary potential \cite{Fra11}
(here considered to be negligible), also leads to an energy relation 
$cp_{b} = \sqrt{E^{2} - (m_{b}c^{2})^{2}}$
as in Eq.(\ref{eq:Ein}) and with $b$ the flavor index.
Thus, the simple above result suggests how a subatomic particle with non-zero mass 
may attain superluminal equilibrium velocities by travelling in helical motion in 
the limit of very large $p$ and small $\theta > 1^{o}$ or $0.01745 \; radians$.  

In our approach, the ratio $v/c$ may label the {\it 'degree of spiralicity'} derived 
from an induced centrifugal force 
along a curved path, or it may denote a sort of 'averaged angle of incidence' 
between two helix points crossed in the least time in analogy with
Fermat's principle.  Our hope with this scenario is to shed new light for
the propagation of superluminal neutrinos without the need to violate 
fundamental laws and well-established astrophysical observations \cite{Gon11}.
As in \cite{Keh11}, the local effect could be caused by a scalar field (here
associated with the centrifugal force of Eq.(\ref{eq:Cen})), which could be 
sourced by distorted regions of spacetime in the surrounding 
environment \cite{Can07}, at least for baseline distances of the 
order of the earth radius.

\end{document}